\documentclass[11pt]{iopart}

\usepackage{iopams}  
\usepackage{url}
\usepackage{subfigure}
\usepackage{graphicx}
\usepackage{xcolor}
\usepackage[normalem]{ulem}
\begin{document}

\title[MoCo-SToRM]{Dynamic imaging using Motion-Compensated SmooThness Regularization on Manifolds (MoCo-SToRM)}

\author{Qing Zou$^1$, Luis A. Torres$^2$, Sean B. Fain$^3$, Nara S. Higano$^{4,5}$, Alister J. Bates$^{4,5}$, Mathews Jacob$^1$}

\address{$^1$Department of Electrical and Computer Engineering, The University of Iowa, Iowa City, IA, USA\\
$^2$Department of Medical Physics, University of Wisconsin, Madison, WI, USA\\
$^3$Department of Radiology , The University of Iowa, Iowa City, IA, USA\\
$^4$Center for Pulmonary Imaging Research, Division of Pulmonary Medicine and Department of Radiology, Cincinnati Children’s Hospital, Cincinnati, OH, USA\\
$^5$Department of Pediatrics, University of Cincinnati, Cincinnati, OH, USA}
\ead{zou-qing@uiowa.edu}
\vspace{10pt}

\begin{abstract}
We introduce an unsupervised motion-compensated reconstruction scheme for high-resolution free-breathing pulmonary MRI. 
We model the image frames in the time series as the deformed version of the 3D template image volume. We assume the deformation maps to be points on a smooth manifold in high-dimensional space. Specifically, we model the deformation map at each time instant as the output of a CNN-based generator that has the same weight for all time-frames, driven by a low-dimensional latent vector. The time series of latent vectors account for the dynamics in the dataset, including respiratory motion and bulk motion. The template image volume, the parameters of the generator, and the latent vectors are learned directly from the k-t space data in an unsupervised fashion. Our experimental results show improved reconstructions compared to state-of-the-art methods, especially in the context of bulk motion during the scans.
\end{abstract}

\vspace{2pc}
\noindent{\it Keywords}: Pulmonary MRI, Motion-Compensated Reconstruction, CNN
%\maketitle

\section{Introduction}

Magnetic resonance imaging (MRI) is an attractive imaging modality for patient groups that require serial followup because it does not use ionizing radiation. Ultra-short echo-time MRI \cite{bracher2011feasibility} methods are capable of significantly reducing the $T_2^*$ losses, mitigating some of the main challenges associated with lung MRI. However, MRI is a slow imaging modality, which makes it challenging to image moving organs such as the lung. For lung MRI, the respiratory motion can be frozen by breath-holds. However, subjects usually are unable to hold their breath for a long time, which will significantly limit the achievable spatial resolution and coverage. Besides, there are several patient groups (e.g., patients with chronic obstructive pulmonary disease (COPD),  pediatric patients, and neonates) who cannot hold their breath even for a short duration \cite{higano2017retrospective,hahn2017pulmonary,higano2018neonatal}. For these reasons, some of these patients need to be sedated for MRI exams or are often not eligible for MRI scans.

Several gating approaches were introduced to eliminate the need for breath-holding in pulmonary MRI \cite{gibiino2015free,kumar2017feasibility,jiang2018motion,capaldi2018free}. For instance, classical methods (e.g., \cite{berecova2012complex}) rely on respiratory bellows or self-gating signals to bin the data to specific phases. Prospective methods only acquire the data during a specific respiratory phase, while retrospective methods continuously acquire the data but only use the data from a specific phase. Self-gating approaches such as XD-GRASP \cite{feng2016xd} use the information from the central k-space samples to estimate the motion signal, which is used to bin the acquired data into several respiratory phases. After the binning, a compressed-sensing approach is used to jointly reconstruct the phase images (See Fig. 1 (a) for illustration), which is more data efficient than traditional binned acquisitions. These approaches are often called motion-resolved methods. A challenge with these methods is the potential sensitivity to bulk motion during the scan. In particular, the subjects may move abruptly during the scan. Because XD-GRASP and similar gating methods rely on low-pass filtering to estimate the pseudo-periodic motion signal, the bulk motion effects are often filtered out. In addition, these approaches are only able to recover the respiratory phase images, which correspond to the averaged data over several minutes, and not the true dynamics. Another challenge associated with the motion-resolved scheme is the trade-off between residual aliasing and blurring resulting from intra-bin motion. For instance, increasing the number of bins can reduce intra-bin motion artifacts. However, this will come at the expense of k-t space data available for each bin, which will translate to residual alias artifacts. While manifold approaches \cite{nakarmi2017kernel,ahmed2020free,zou2021dynamic,yoo2021time}, which perform soft-gating as opposed to explicit binning, offer improved trade-offs but are also vulnerable to these challenges. These schemes use machine learning algorithms to perform soft-binning of the data using the manifold structure of images in the dataset. These unsupervised machine learning methods have been shown to offer improved performance and robustness to different motion patterns over explicit binning strategies.

Motion compensation (MoCo) is often used to further improve the data efficiency and to reduce residual aliasing and noise in the reconstructed images. Many of the approaches require a high-resolution reference image. The recovered images are then registered to the reference image to obtain the motion fields \cite{huttinga2021nonrigid}. Another approach is to estimate the motion-maps between the phase images reconstructed by XD-GRASP; the different motion phases are registered together and averaged to obtain a MoCo volume \cite{jiang2018motion}. Recently, Zhu et al. used a non-linear compressed sensing algorithm to directly recover the MoCo volume from the k-t space data, using the deformation maps estimated from XD-GRASP \cite{zhu2020iterative}. This approach, named iMoCo, is shown to significantly improve the image quality. The main challenge with this multi-step strategy (binning based on motion estimation, followed by XD-GRASP reconstruction, followed by the final MoCo reconstruction) is the dependence of the image quality on the intermediate steps. In particular, this approach inherits the sensitivity of XD-GRASP to bulk motion because it is dependent on motion estimates from XD-GRASP. For example, when the data is corrupted with a single bulk motion effect, the data with bulk motion is removed in \cite{zhu2020iterative} to obtain good reconstructions; this approach is not readily applicable to settings where there are multiple bulk motion events during the acquisition. This multi-step strategy was recently extended to dynamic PET \cite{li2020motion}. The main distinction of this scheme from \cite{zhu2020iterative} is the use of a deep learning algorithm to estimate the deformation maps between the different motion phases. The different motion phases are registered to a fixed state using a deep network, followed by a reconstruction scheme similar to iMoCo \cite{zhu2020iterative}. 

The main focus of this work is to introduce a novel unsupervised deep-learning MoCo reconstruction scheme, which can be readily applied for free-breathing pulmonary MRI. This method is the generalization of the previous motion-resolved generative manifold methods \cite{zou2021dynamic,yoo2021time} to the MoCo setting; we hence call the proposed approach motion-compensated smoothness regularization on manifolds (MoCo-SToRM). Unlike \cite{zou2021dynamic,yoo2021time}, which assume the images to be on a smooth manifold, we assume that the motion deformation maps at different time instants are living on a manifold, parameterized by low-dimensional latent vectors. We assume the deformation maps to be the output of a convolutional neural network (CNN) based generator, whose inputs are time-dependent low-dimensional latent vectors that capture the motion information (See Fig. 1 (b) for illustration). The generated deformation maps are used to deform a learned template image, which corresponds to the image volume frame in the time series. A multi-channel non-uniform Fourier transform (NUFFT) is used to generate the k-space measurements of the images. Unlike prior MoCo approaches that use a series of algorithms for binning, reconstruction, motion estimation, and reconstruction, we formulate the joint recovery of the latent vectors, deformation maps, and the template image directly from the measured k-t space data as a single non-linear optimization scheme. The cost function is the squared error between the multi-channel Fourier measurements of the image volumes and the actual measurements acquired from the specific subject. We note that the deformation maps are smooth and are less complex than the images themselves; we expect the proposed scheme to be less data-demanding than motion-resolved approaches \cite{feng2016xd,zou2021dynamic,yoo2021time}. 

The proposed framework, built using modular blocks, is highly explainable. In particular, the learned latent vectors capture the intrinsic temporal variability in the time series, including respiratory and bulk motion as seen from our experiments. Moreover, the smooth deformation maps capture the spatial deformation of the image template and can be visualized. More importantly, the reconstructions can be viewed as a movie, allowing one to visualize the images at respective time-frames, unlike binning-based approaches that only recover the phase images. Unlike current deep-learning strategies that pre-learn the CNN parameters from example data, the proposed scheme learns all the parameters from the data of the specific subject. We also note that the MoCo approach enables us to minimize the trade-off between intra-phase motion and the data available for reconstruction. We do not make any assumptions on the latent vectors, which allows it to learn all the motion events during the acquisition, including bulk motion, which is challenging for traditional methods. 

\section{Methods}

\subsection{Brief background on motion-resolved and motion-compensated reconstruction}
\begin{figure}[!thb]
\centering
           \subfigure[Motion-resolved reconstruction (XD-GRASP)]{\includegraphics[width=0.6\textwidth]{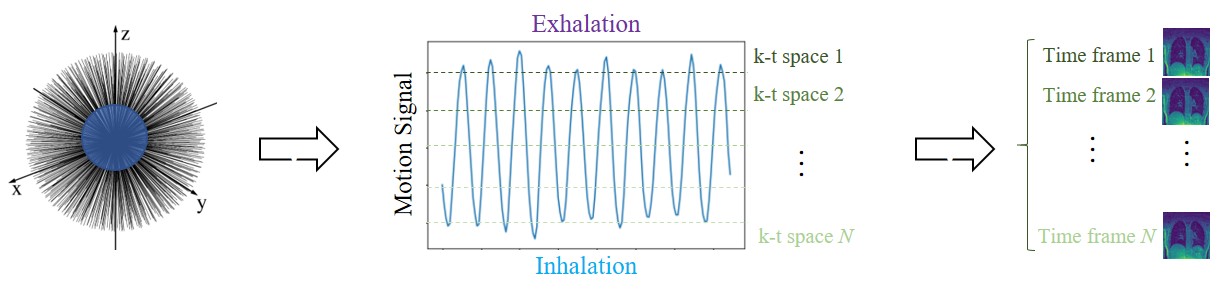}}\vspace{2em}
	\subfigure[Proposed MoCo-SToRM]{\includegraphics[width=0.9\textwidth]{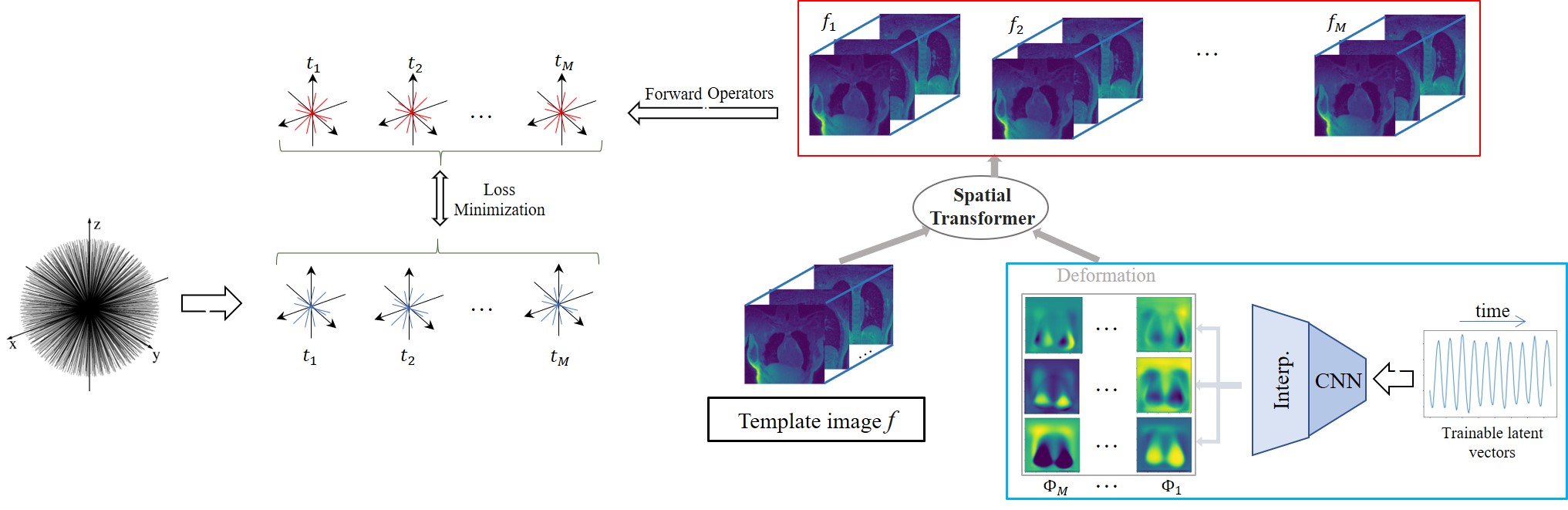}}
	\caption{\small{Illustration of XD-GRASP and MoCo-SToRM algorithms. (a) XD-GRASP uses the central k-t space data to estimate the motion signals and bins the k-t space data into several motion phases. The binned data are then used for the recovery of the images using total variation and wavelet regularization. (b) The proposed MoCo-SToRM scheme jointly learns the deformation maps $\boldsymbol{\phi_t}$ and the static image template $f$ from the k-t space data. To regularize the deformation maps, we model $\phi_t = \mathcal G_{\theta}(\mathbf{z}_t)$ as the outputs of a deep CNN generator $\mathcal G_{\theta}$ whose weights are denoted by $\theta$. The inputs of the generator are the low-dimensional (e.g., 1-D in lung MRI) latent vectors. The parameters of the CNN generator $\theta$, the latent vectors $\mathbf{z}_t$, and the template $f$ are jointly estimated from the data. The loss is the mean square error between the actual measurements and the multi-channel measurements of the deformed images $f_t$.}}
	\label{illu}
\end{figure}

Several self-navigated motion-resolved free-breathing MRI schemes \cite{feng2016xd,feng2019simultaneous}, which use 3D radial ultra-short-echo (UTE) acquisition, were recently introduced for lung imaging. These schemes rely on a combination of low-pass filtering and/or clustering to derive the self-gating signals, which are used to bin the k-space data into different motion phases. Once the data is binned, these schemes (e.g., XD-GRASP \cite{feng2016xd}) perform the motion-resolved joint reconstruction of the phases by solving the following:
\[\mathbf F^* = \arg\min_{\mathbf F}||\mathcal{A}(\mathbf F)-\mathbf{B}||_2^2 + \lambda_s||\Psi\mathbf F||_1+\lambda_t {\rm TV}(\mathbf F).\]
The first term is the data consistency term that compares the multi-channel measurements of the phase images $\mathbf F = \{f_1,\cdots,f_N\}$ with the binned data $\mathbf B = \{\mathbf b_1,\cdots,\mathbf b_N\}$. Here $N$ is the number of bins and hence the number phases in the XD-GRASP reconstruction. The second term is a spatial sparsity $\ell_1$-wavelet penalty term, in which $\Psi$ is the wavelet transform. The third term is the total variation penalty along the motion phases. The above scheme is usually called XD-GRASP-type motion-resolved reconstruction, which is illustrated in Fig. \ref{illu} (a).

Deep manifold-based approaches \cite{zou2021dynamic,zou2021deep} offer an alternate route for motion-resolved recovery. In particular, the images in the time series are modeled as $f_t = \mathcal G_{\theta}(\mathbf z_t)$, where $\mathcal G_{\theta}$ is a deep CNN generator that is shared across different image frames. The parameters of the generator denoted by $\theta$ and the low-dimensional latent vectors $\mathbf z_t$ are learned such that the cost function $\sum_t \|\mathcal A_t \mathcal G_{\theta}(\mathbf z_t) - \mathbf b_t\|^2$ is minimized. Here, $\mathcal A_t$ denotes the forward model corresponding to the $t^{\rm th}$ image frame, and $\mathbf b_t$ are the corresponding k-space measurements. The proposed approach is a generalization of these deep manifold models \cite{zou2021dynamic,zou2021deep} to the MoCo setting. 

Many of the early MoCo methods rely on a high-resolution static reference image \cite{huttinga2021nonrigid}. Recent approaches \cite{jiang2018motion,zhu2020iterative} rely on motion-resolved XD-GRASP reconstructions. Once the motion-resolved reconstructions are obtained, one of the phases (usually the exhalation phase) will be chosen as the reference. Then the other motion phases are registered to the reference phase to obtain the deformations $\Phi_1,.. \Phi_p$. The iMoCo approach solves the following optimization scheme to obtain the MoCo reconstruction $f_{\rm imoco}$:
\[f_{\rm imoco}^* = \arg\min_{f_{\rm imoco}}\sum_{p=1}^{N} ||\mathcal{A}_t(\Phi_p(f_{\rm imoco})) - \mathbf{b}_p||_2^2 + \lambda {\rm TGV}(f_{\rm imoco}).\]
The first term is the data consistency term, and the second is a spatial total generalized variation sparsity regularizer \cite{knoll2011second}. 

\subsection{Proposed approach}
In this work, we extend the motion-resolved deep manifold methods \cite{zou2021dynamic,zou2021deep} to the MoCo reconstruction setting. The proposed framework is a end-to-end self-supervised deep-learning algorithm involving explainable learning modules. We model the image volumes $f_t$ in the time series as the deformed versions of a single image template $f$:
\begin{equation}\label{recon1}
f_t(x,y,z) = f\Big(x-\phi_{x}(t),y-\phi_{y}(t),z-\phi_{z}(t)\Big) :=  \mathcal{D}\left({f},\boldsymbol\phi(t)\right).
\end{equation}
Here $\boldsymbol \phi(t) = \{\phi_{x}(t), \phi_{y}(t), \phi_{z}(t)\}$ is the motion/deformation map at the time instant $t$. We implement $\mathcal{D}$ as a differentiable interpolation layer. 

We propose to jointly estimate the deformation maps and the single image template $f$ directly from the k-t space data. We note that it is impossible for us to acquire the k-space data at the Nyquist–Shannon sampling rate. Therefore, the joint estimation problem is highly ill-posed. In order to regularize the deformation maps, we use the manifold assumption. In other words, we assume that the deformation map for each image frame $f_t$ is living on a smooth manifold, parameterized by low-dimensional latent vectors $\mathbf{z}_t,\,\, t = 1,\cdots M$ that capture the dynamics (e.g., respiratory motion, bulk motion in lung imaging). We model the non-linear mapping between the low-dimensional latent vectors and the high-dimensional deformation maps by a CNN generator:
 \begin{equation}\label{generator}
     \boldsymbol \phi(t)= \mathcal{G}_{\theta}\left(\mathbf{z}_t\right),
 \end{equation}
 whose input is the low-dimensional latent vector $\mathbf z_t \in \mathbb R^d$. As the dominant motion in free-breathing lung imaging is the respiratory motion, we set $d=1$ in this work; we will consider higher dimensional latent space in our future work. Combining (\ref{recon1}) and (\ref{generator}), each image frame $f_t$ in the time series is modeled as:
\begin{equation}\label{recon}
f_t(\mathbf r) = \mathcal{D}\left({f},\underbrace{\mathcal{G}_{\theta}(\mathbf{z}_t)}_{\boldsymbol\phi(t)}\right).
\end{equation}
Here, $\mathbf r = (x,y,z)$ is the spatial coordinate. See Fig. \ref{illu}.(b) for an illustration. 

We note that the generated image at each time instant $t$ is dependent on the image template $f$, the parameters of the deep CNN generator $\theta$, and the low-dimensional latent vectors $\mathbf z = [\mathbf z_1,..,\mathbf z_M]$. Here, $M$ is the number of image frames in the time series. When golden-angle or bit-reversed radial acquisitions are used, $M$ can be a user-defined parameter. We propose to jointly solve for the above unknowns directly from the k-t space data of the specific subjects as the optimization problem:
\begin{equation}\label{cost}
\mathcal{C}(\mathbf{z},\theta,{f}) = \sum_{t=1}^M||\mathcal{A}_t({f}_t)-\mathbf{b}_t||^2 + \lambda_{1}|\nabla_t\mathbf{z}\|_{\ell_1} + \lambda_2 \|\nabla_{\mathbf r} f\|_{\ell_1},
\end{equation}
where ${f}_t$ is related to the static image $f$ by (\ref{recon}). Here, $\mathcal{A}_t$ are the forward operators that are performed on each of the time points. We implement $\mathcal A_t$ as multi-channel NUFFT \cite{fessler2003nonuniform} operators using the k-space trajectory at the $t^{\rm th}$ time instant using the SigPy package \cite{ong2019sigpy}. $\mathbf{b}_t$ are the multi-channel k-space measurements acquired from the subject. The second term in (\ref{cost}) is a smoothness penalty on the latent vector that captures the dynamics (e.g., respiratory and/or bulk motion). If this term is not added, the learned latent vectors will learn high-frequency oscillations. To minimize this risk, we added a total-variation penalty on the latent vector $\mathbf{z}$ along the time direction to encourage the latent vectors to learn piecewise smooth motion. The last term in (\ref{cost}) is the spatial total variation penalty on the static image, which enables us to further reduce alias artifacts in the learned static image. 

The proposed self-supervised scheme offers several benefits. First of all, the reconstruction relies only on the undersampled data acquired from the specific subject. Unlike most deep-learning strategies, the proposed framework does not require fully sampled training datasets, which are not available in our setting, to train the networks. Secondly, the proposed scheme does not require physiological monitors such as respiratory belts or dedicated k-space navigators. It also eliminates the need for band-pass filtering or clustering to estimate the phase information, which will filter out bulk motion. Finally, unlike binned approaches that recover average images over the acquisition duration within respective respiratory phases, the proposed scheme enables the recovery of the natural dynamics of the lung.

\subsection{Approaches to minimize computational complexity}
 We use ADAM optimization \cite{adam2015} with a batch size of one time-frame to find the optimal $\mathbf{z},\theta$ and $f$. The small memory footprint enables us to use this scheme for high-resolution 3D+time problems. The network and optimization scheme was implemented in PyTorch. The motion generator is implemented using an eight-layer network. The first seven layers are 3D convolutional layers with 200 features per layer. The last layer is an up-sampling layer, which uses tri-linear interpolation to interpolate the deformation maps from lower resolution to high resolution. The final interpolation step allows us to account for the prior knowledge that the deformation maps are smooth functions. ReLU activation function \cite{agarap2018deep} is used for all the convolutional layers.

Directly solving the above optimization problem (\ref{cost}) is computationally expensive, especially when the image resolution is high. To minimize the computational complexity, we use a progressive strategy. In particular, we first solve for (\ref{cost}) for very low-resolution images using the corresponding region in the central k-t space. These latent vectors, motion fields, and the images are used to initialize the network at a higher spatial resolution. We use two progressive steps to refine the resolution, until the final resolution is reached. We observe that this progressive strategy significantly improves the convergence rate. In particular, few iterations are needed at the highest resolution, compared to the setting where the parameters of the motion network and the image are initialized randomly. 

The above joint optimization strategy offers good estimates of the latent vectors with few iterations, even at the lowest resolution. By contrast, the stochastic gradients with a batch size of one can result in low convergence rates for the static image $f$ and the CNN parameters. To further accelerate the convergence rate, we additionally use a binning strategy similar to motion-resolved schemes shown in Fig. \ref{illu}.(a), assuming $\mathbf z$ to be fixed. We bin the latent vector to $P$ phases, based on the latent vectors we estimated. We use 25 phases in the adult subjects with less extensive motion and 150 in the neonatal intensive care unit (NICU) patient with extensive motion. This approach allows us to bin the data to different phases. We update $\theta$ and $f$ using the optimization strategy:
\begin{equation}\label{binned}
\mathcal{C}(\theta,{f}) = \sum_{p=1}^P\left\|\mathcal{A}_p\underbrace{\Big(\mathcal D\left(f, \mathcal G_{\theta}(\mathbf{z}_p)\right)\Big)}_{f_p}-\mathbf{b}_p\right\|^2 + \lambda_2 \|\nabla_{\mathbf r} f\|_{\ell_1},
\end{equation}
Here, $\mathbf{z}_p$ is the latent vector at the $p^{\rm th}$ bin, and $\phi_p=\mathcal G_{\theta}(z_p)$ is the motion vector for the $p^{\rm th}$ bin. Here $f_p$ is the image in the $p^{\rm th}$ bin, obtained by deforming $f$ with $\phi_p$. 

We only use the above binning-based optimization in (\ref{binned}) at the highest resolution level. Note that (\ref{binned}) only solves for $\theta$ and $f$, and not $\mathbf z$. We hence alternate between (\ref{binned}) and (\ref{cost}) at the highest resolution.

\section{Datasets and evaluation}

\subsection{Experimental datasets}

The datasets used in the experiments in this work were acquired using an optimized 3D UTE sequence with variable-density readouts to oversample the k-space center \cite{johnson2013optimized}. We used four datasets acquired from two adult subjects. Two of them are from a healthy subject (pre-contrast and post-contrast) and another two are from a fibrotic subject (pre-contrast and post-contrast). We also used one dataset acquired from a female subject with severe bronchopulmonary dysplasia (BPD), who was admitted to the NICU. The gestational age of the patient at birth is 24 weeks, and the MRI is prescribed at the chronological age of 15 weeks and 3 days. The weight of the patient at MRI is 3.18 Kg.

The datasets from the healthy subject were acquired on a 1.5T GE scanner using 8 coils. The variable-density readouts help retain signal-to-noise ratio (SNR), and oversampling reduces aliasing artifacts. A bit-reversed ordering was used during the data acquisition. The prescribed field of view (FOV) $= 32 \times 32\times 32$ cm$^3$. The matrix size is $256\times 256\times 256$. The data were acquired with 90K radial spokes with TR$\approx$ 3.2 ms and 655 samples/readout, corresponding to an approximately five-minute acquisition.

The datasets from the diseased subject were acquired on a 3T GE scanner using 32 coils. The prescribed FOV $= 32 \times 32\times 32$ cm$^3$. The matrix size is $256\times 256\times 256$. The data were acquired with 91K radial spokes with TR$\approx$ 2.8 ms and 655 samples/readout, corresponding to an approximately four-minute acquisition. When we were processing the datasets with 32 coils, we used a PCA-based coil combination  \cite{pedersen2009k} using SVD to keep only eight virtual coils.

The dataset from the neonatal subject was acquired on a 1.5T small footprint MRI scanner \cite{hahn2017pulmonary} located in the NICU. The data was acquired using the built-in body coil, which translates to poor SNR compared to the adult scans. In addition, the scan is made challenging because of the extensive bulk motion by the subject during the scan. For this dataset, the prescribed FOV $= 18 \times 18\times 18$ cm$^3$. The matrix size is $256\times 256\times 256$. The data was acquired with 200K radial spokes with TR$\approx$ 5 ms and 1013 samples/readout, corresponding to an approximately 16-minute acquisition.

This research study was conducted using human subject data. Approval was granted by the Ethics Committees of the institutions where the data was acquired.

\subsection{Numerical Phantom to validate MoCo-SToRM}

\begin{figure}[!htbp ]
\centering
           \includegraphics[width=0.5\textwidth]{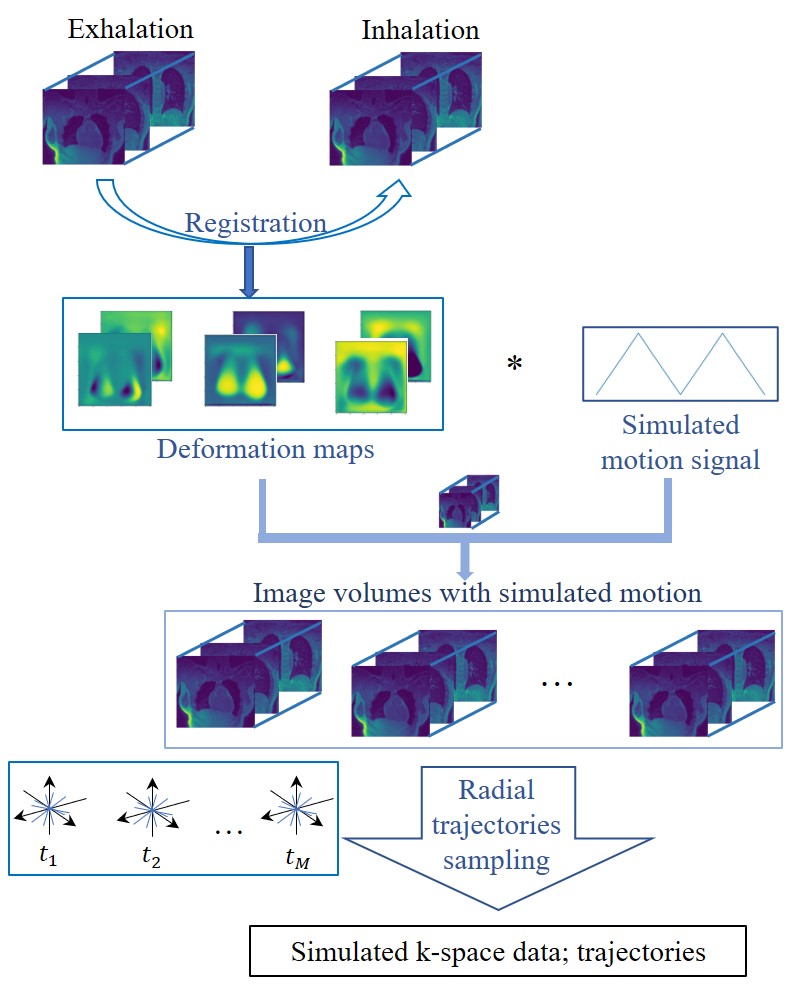}
	\caption{\small{Illustration of the numerical Phantom data generation. We create the simulation data using the reconstructions from XD-GRASP and iMoCo following the process shown in the figure.}}
	\label{data_gen}
\end{figure}

We note that we do not have accurate ground truth datasets to evaluate the quantitative accuracy of the proposed scheme and its potential impact to bulk motion. We hence constructed a high-resolution numerical phantom using the XD-GRASP and iMoCo reconstructions of the pre-contrast dataset from a healthy subject. Specifically, we registered the XD-GRASP exhalation phase to the inhalation phase to obtain the deformation maps. Then we modulated the deformation maps by a periodic triangular function with a specific frequency and a DC off-set to simulate the motion from the exhalation phase to the inhalation phase. To study the impact of bulk motion, we also consider additional random translational motion (move about $2$ - $5$ pixels), at random time points as shown in Fig. \ref{sim_study}. We deform the static iMoCo reconstruction with the above deformation maps to generate the high-resolution images at different time points. The multi-channel NUFFT of these images were used as the measurements, with an additive white Gaussian noise of $0.5\%$. The creation of the simulation data is illustrated in Fig. \ref{data_gen}.

\subsection{Figures of merit for quantitative evaluation}

For image quality comparisons, we compare the proposed MoCo-SToRM reconstruction with XD-GRASP and iMoCo. To quantitatively compare the image quality, we use three images metrics in this work.

\begin{itemize}
\item  Diaphragm maximum derivative (DMD) \cite{zhu2020iterative,breuer2018stable}: the DMD will be used to measure the sharpness of the lung-liver diaphragm. It is defined as:
\[{\mathrm{DMD}} = \frac{{\mathrm{Max}}({\partial{I}})}{{\mathrm{Mean}}(I_{liver})},\]
where ${\mathrm{Max}}({\partial{I}})$ is the maximum intensity change between the lung–liver interface, which is computed by choosing the maximum value of the image gradient. ${\mathrm{Mean}}(I_{liver})$ is the mean intensity in the chosen liver region. A higher DMD implies sharper edges.

\item Signal-to-noise ratio (SNR) \cite{dietrich2007measurement,gutberlet2018free,kobayashi2021magnetic}: The SNR is computed as
\[{\mathrm{SNR}} = 20\log(\frac{\mu_s}{\sigma_n}),\]
where $\mu_s$ is the mean of the intensity of the chosen region of interest and $\sigma_n$ is the standard deviation of the intensity of a chosen noise region. A higher SNR usually means better image quality. In our study, we manually choose the regions of interest.

\item Contrast-to-noise ratio (CNR) \cite{zhu2020iterative,prince2006medical}: The CNR is computed as
\[{\mathrm{CNR}} = 20\log(\frac{|\mu_A-\mu_B|}{\sigma_n}),\]
where $\mu_A$ and $\mu_B$ are the mean of the intensity of two regions within the region of interest and $\sigma_n$ is the standard deviation of the intensity of a chosen noise region. The higher CNR usually means better image quality. In our study, we manually choose the regions of interest.
\end{itemize}

\section{Results}

\subsection{Numerical simulation experiments}

\begin{figure}[!htbp ]
\centering
    \subfigure[Quantitative results]{\includegraphics[width=0.55\textwidth]{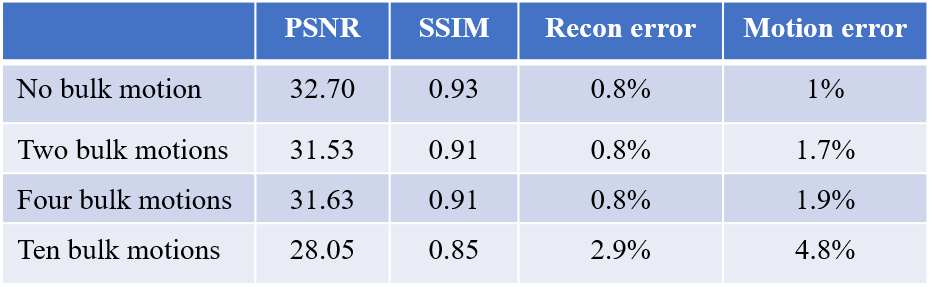}}\\
	\subfigure[Simulation results]{\includegraphics[width=0.85\textwidth]{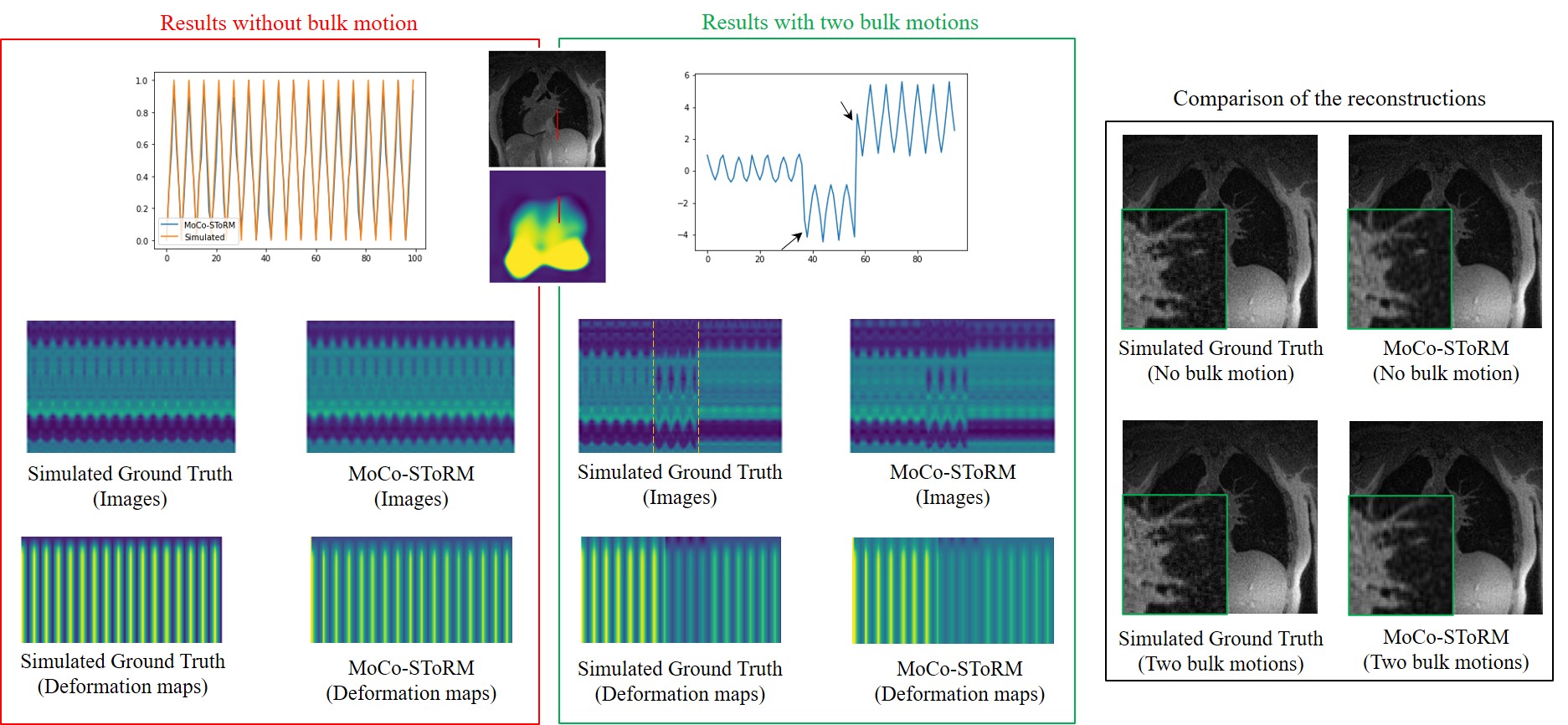}}\\
	\subfigure[Simulation results]{\includegraphics[width=0.85\textwidth]{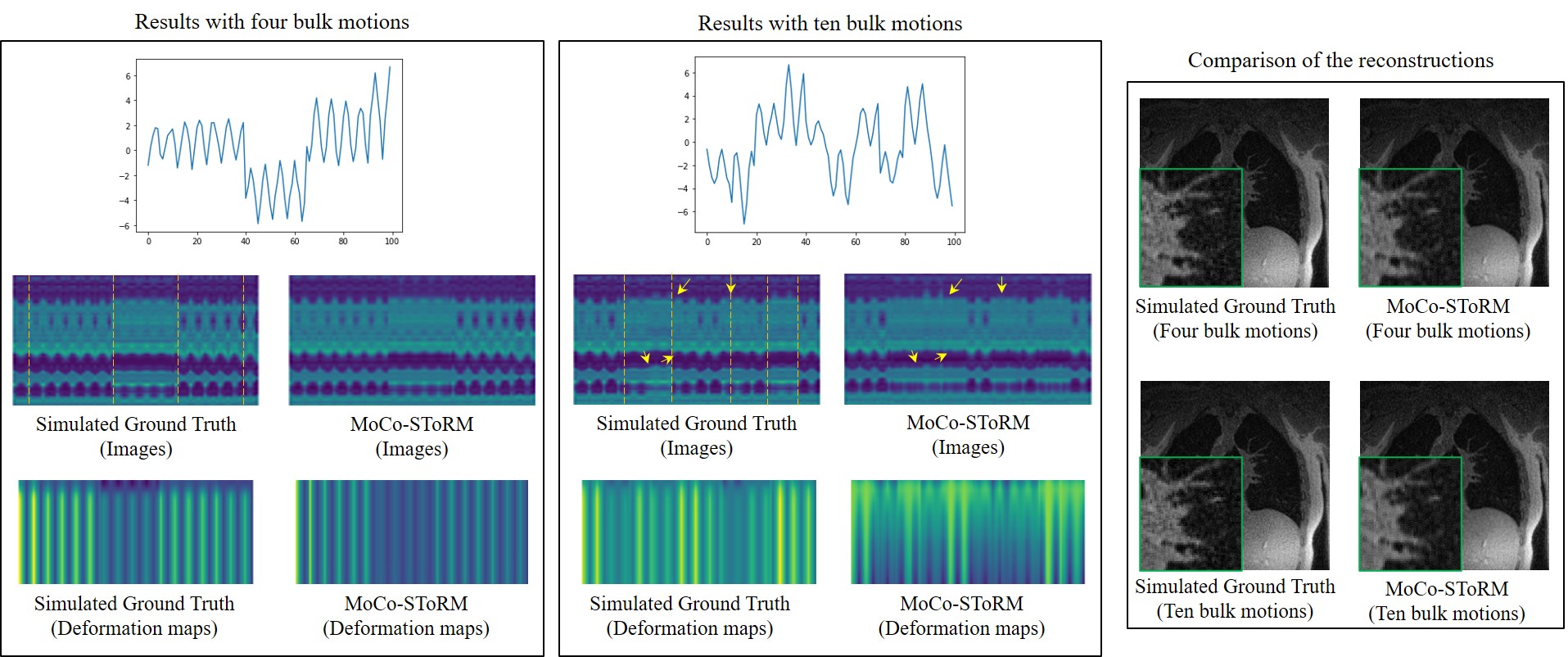}}
	\caption{\small{Results of the numerical simulation study. The proposed MoCo-SToRM scheme is applied on the simulation data, where we simulated the data with no, two, four, and ten bulk motions. The quantitative results of the simulation are shown in (a). In (b) and (c), we show the learned latent vectors, the time profiles of the image volumes, the time profiles of the deformation maps, and the comparison of the reconstructions for the four cases. We note that the proposed scheme can offer reliable estimates when there are few bulk motion events. When the number of bulk motion events increases, the amount of k-t space data available in each bulk motion state decreases. The reduced data translates to higher motion errors, which results in increased reconstruction error.  }}
	\label{sim_study}
\end{figure}

The results of the simulation study are shown in Fig. \ref{sim_study}. In this simulation study, we investigate the impact of bulk motion events. Specifically, we create the simulation data with no, two, four, and ten bulk motion events. We quantitatively compare the reconstructions using the metrics of PSNR, SSIM, relative error of the reconstruction, and relative error of the deformation maps. The quantitative results of the simulation study are summarized in Fig. \ref{sim_study} (a). In (b) and (c), we show some results of the simulation study. We show the learned latent vectors, the time profiles of the reconstructed image volumes, and the deformation maps. The comparisons of the reconstructions are also shown in the figure. From the simulation study, we see that the proposed MoCo-SToRM approach works reasonably when there are four bulk motion events. By contrast, when there are ten bulk motion events, the performance of the proposed scheme degrades in a graceful fashion. In particular, the limited number of radial k-space spokes translates to imperfect estimation of motion, which in turn translates to blurry reconstructions.

\subsection{Experimental datasets}

\begin{figure}[!htbp ]
\centering
           \includegraphics[width=0.95\textwidth]{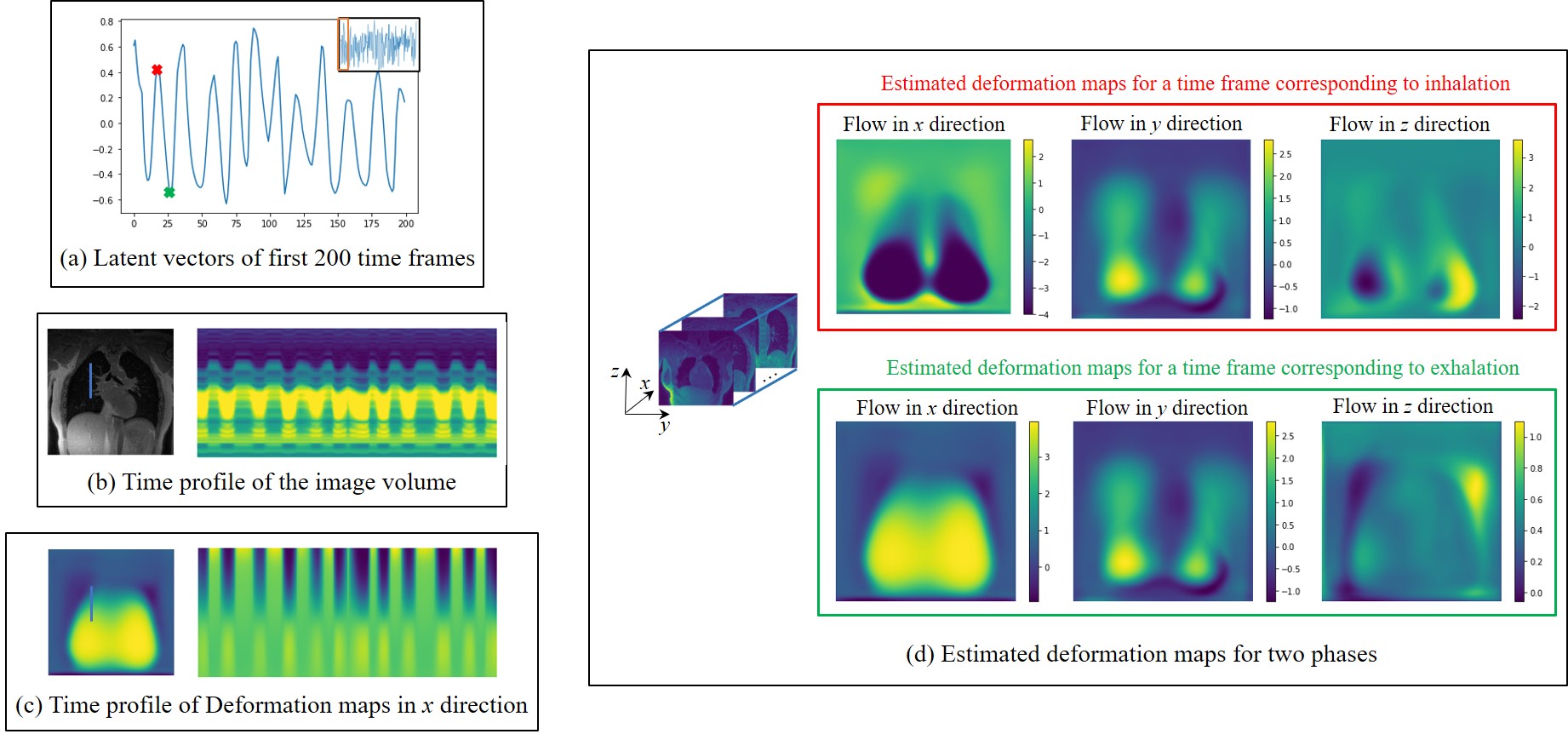}
	\caption{\small{Illustration of the learned quantities from the pre-contrast healthy volunteer. In (a), we show the estimated latent vectors corresponding to the first 200 frames, while (b) and (c) show the time profile of the reconstructed image volumes and the deformation maps, respectively. From the three figures on the left, we see that the motion patterns in the time profiles closely match the learned latent vectors. In (d), we show the deformation maps in the three directions, corresponding to the time frames marked by red and green cross marks in the latent vectors in (a).}}
	\label{lat_flow}
\end{figure}

In Fig. \ref{lat_flow}, we show the learned latent vectors, time profiles, and example estimated deformation maps and corresponding time profiles from the pre-contrast dataset from the healthy subject. We show the estimated latent vectors from the first 200 frames in (a). We also show the time profile of the reconstructed images in (b) and the time profile of the deformation maps in (c), corresponding to the blue lines in the images. From the two profiles, we see that the motion patterns coincide with the learned latent vectors. In (d) we show the estimated deformation maps from two time points, indicated by red and green dots in (a), corresponding to the inhalation phase and the exhalation phase. The results show that the latent vectors closely capture the dynamics of the motion.

\subsection{Comparison with state-of-the-art methods}
\begin{figure}[!thb]
\centering
    \subfigure[Visual comparison (healthy subject)]{\includegraphics[width=0.6\textwidth]{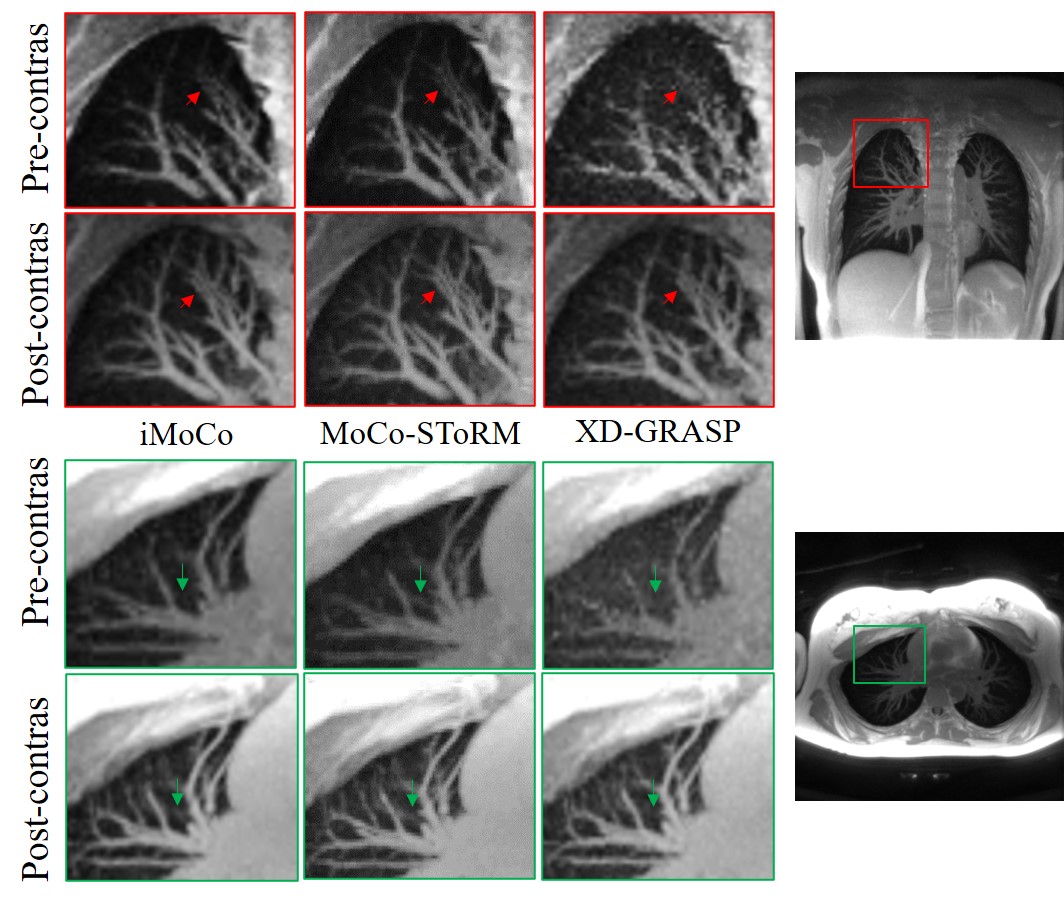}}\\
	\subfigure[Visual comparison (diseased subject)]{\includegraphics[width=0.6\textwidth]{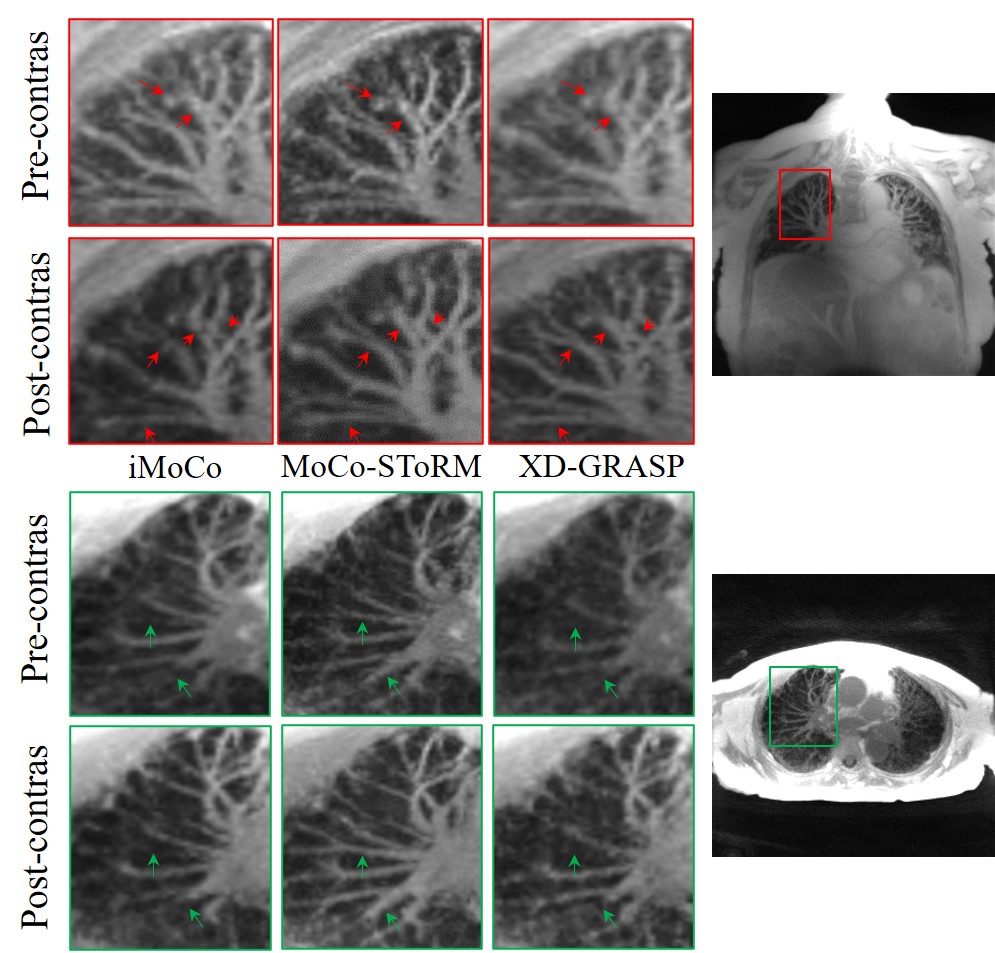}}
	\caption{\small{Visual comparison of the reconstructions from different methods on two post-contrast datasets. In (a), we show the results from the healthy subject. Two regions from both the sagittal view are shown in the figure. In (b), we show the results obtained from the diseased subject. From the figure, we can see that the proposed MoCo-SToRM scheme is able to reduce the noise and blur when compared to the competing methods.}}
	\label{vis_comp}
\end{figure}

In this section, we compare the results of the proposed scheme with XD-GRASP and iMoCo. In Fig. \ref{vis_comp}, we show the visual comparisons of the methods on post-contrast datasets. From Fig. \ref{vis_comp}, we observe that the MoCo-SToRM reconstructions can reduce the noise and capture more details when compared to the motion-resolved XD-GRASP reconstructions. Furthermore, the MoCo-SToRM reconstructions are less blurred than those of the motion-compensated iMoCo reconstructions. We note that the post-contrast dataset from the diseased subject had a bulk motion event (see Fig. \ref{bulk_motion}), which translates to blurred iMoCo reconstructions.

\begin{figure}[!thb]
\centering
      \subfigure[DMD comparison]{\includegraphics[width=0.5\textwidth]{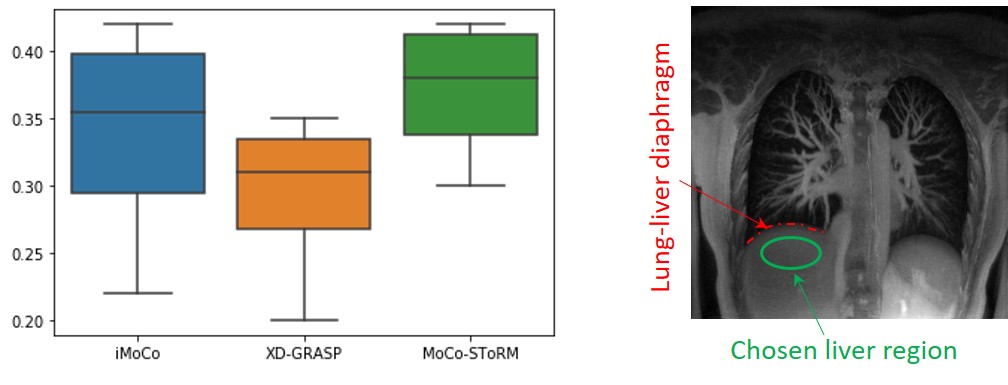}}
      \subfigure[SNR and CNR comparison]{\includegraphics[width=0.7\textwidth]{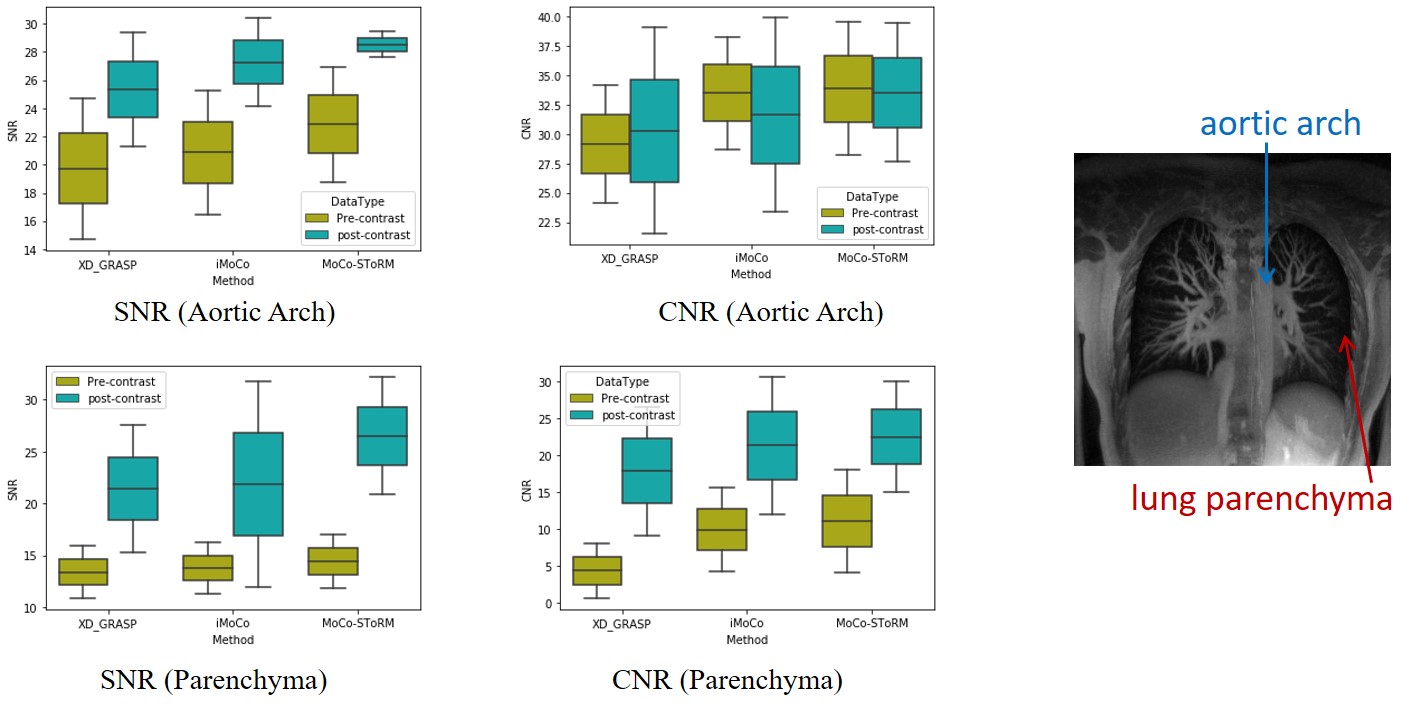}}
	\caption{\small{Results of quantitative comparison. (a) shows the DMD results, from which we can see that the proposed MoCo-SToRM scheme is able to provide comparable DMD results to those of the iMoCo scheme. (b) shows the results of the SNR and CNR comparison. We can see from the figure that the proposed MoCo-SToRM results are comparable with those obtained from the motion-compensated iMoCo scheme.}}
	\label{snr_cnr}
\end{figure}

\begin{figure}[!htbp ]
\centering
           \subfigure[Latent vectors and motion detection]{\includegraphics[width=0.45\textwidth]{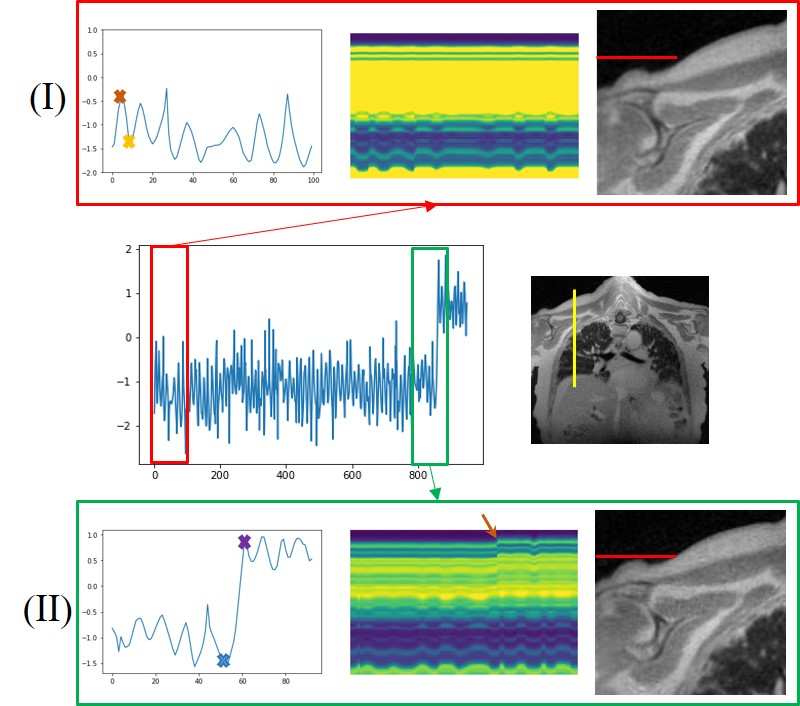}}\hspace{2em}
	\subfigure[Exemplar deformation maps]{\includegraphics[width=0.45\textwidth]{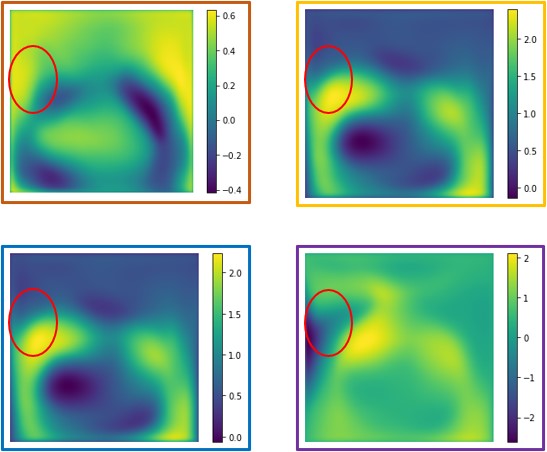}}
	\caption{\small{Impact of bulk motion. (a) shows the latent vectors estimated from the proposed scheme, which are zoomed to regions without bulk motion (I) and regions with bulk motion (II), captured by the discontinuity in the latent vectors. We plot the time profiles at the position marked by the yellow line in the image shown in the middle row. From the plots of the time profiles, we see the subject moved his shoulder during the scan, evidenced by the rightmost reconstructed frame in (I) and (II), respectively; the red lines are in the same location, indicating motion in the shoulder. We also show show four exemplar deformation maps corresponding to four time points marked in (I) and (II). We note that the deformation maps with yellow and blue borders corresponding to the local minima of the latent vectors are similar. By contrast, the local maxima of the map with the purple border shows significant deviation from the one with the red border. }}
	\label{bulk_motion}
\end{figure}

The quantitative comparison of the proposed scheme with the competing methods on four datasets (two from healthy adult subjects and two from diseased adult subjects) are shown in Fig. \ref{snr_cnr}. We first measure the DMD on 15 sagittal slices in each dataset, and the quantitative results are shown in Fig. \ref{snr_cnr} (a). From the DMD results, we see that the proposed MoCo-SToRM scheme is able provide comparable results. In particular, the motion-compensated methods (i-MoCO and MoCO-SToRM) are observed to yield marginally higher DMD than XD-GRASP, implying reduced blurring, as shown in Fig. \ref{snr_cnr} (a).

In addition to DMD, we also report the SNR and CNR of the aortic arch and lung parenchyma region, as shown in Fig. \ref{snr_cnr} (b). We see  that the proposed MoCo-SToRM results are comparable with the results obtained from the motion-compensated iMoCo.

\subsection{Impact of bulk motion}

The acquisition time in pulmonary MRI is around four minutes for adult subjects and around 16 minutes for NICU subjects. The relatively long scan time makes current approaches vulnerable to bulk motion artifacts, especially during the imaging of diseased and pediatric patients. If they are not compensated, these bulk motion errors translate to residual blurring.

In cases with significant bulk motion, existing methods use additional image-based approaches to detect and reject sections of data with bulk motion  \cite{zhu2020iterative}. These approaches are readily applicable to cases with very few bulk motion events; for instance, a case with a single bulk motion event was considered in \cite{zhu2020iterative}. If multiple events are in the dataset, this approach may severely restrict the available k-t space data and hence translate to significantly degraded image quality.

An advantage of the proposed MoCo-SToRM scheme is its ability to directly account for bulk motion during the scan. In the proposed MoCo-SToRM scheme, the latent vectors and the CNN-based generator have the ability to capture the bulk motion in the data and account for it during the reconstruction. In particular, we observe sudden jumps in the learned latent vectors. The non-linear nature of the learned generator allows the generation of deformation maps for each motion state, depending on the value of the latent vectors. 

\subsubsection{Adult study:}

In the post-contrast dataset acquired from a diseased subject, we detected one bulk motion, which is shown in Fig. \ref{bulk_motion}. In Fig. \ref{bulk_motion} (a), we show the latent vectors that are learned using MoCo-SToRM. As we mentioned before, the sudden jump in the latent vectors indicates the bulk motion. We highlight two time regions, one without bulk motion (red box) and one with bulk motion (green box). We also show the time profiles of the yellow line indicated in Fig. \ref{bulk_motion} (a). From the plots of the time profiles, we see that when the latent vectors have no sudden jump, then no bulk motion can be seen. However, when sudden jump happens in the latent vectors, we clearly see a bulk motion event from the time profile. In (I) of Fig. \ref{bulk_motion} (a), we zoomed the reconstructed image corresponding to the yellow cross shown in the latent vectors, and in (II) of Fig. \ref{bulk_motion} (a), we zoomed the reconstructed image corresponding to the purple cross shown in the latent vectors. From the red line marker, we can see that the subject moved the shoulder during the scan.

In Fig. \ref{bulk_motion} (b), we also show the deformation maps estimated by the proposed scheme at four different time points indicated by the brown, yellow, blue, and purple crosses in Fig. \ref{bulk_motion} (a). From the red ellipses in the images, we see that the subject moved the shoulder during the scan. The deformation map will be very different from the deformation maps when there is no bulk motion. In Fig. \ref{vis_comp} (b), we compared the reconstructions from the proposed scheme and the iMoCo and XD-GRASP reconstructions. From the figure, we see that the proposed scheme is able to deal with bulk motion and offer improved reconstructions.

\begin{figure}[!htbp ]
\centering
    \subfigure[Latent vectors and motion detection]{\includegraphics[width=0.65\textwidth]{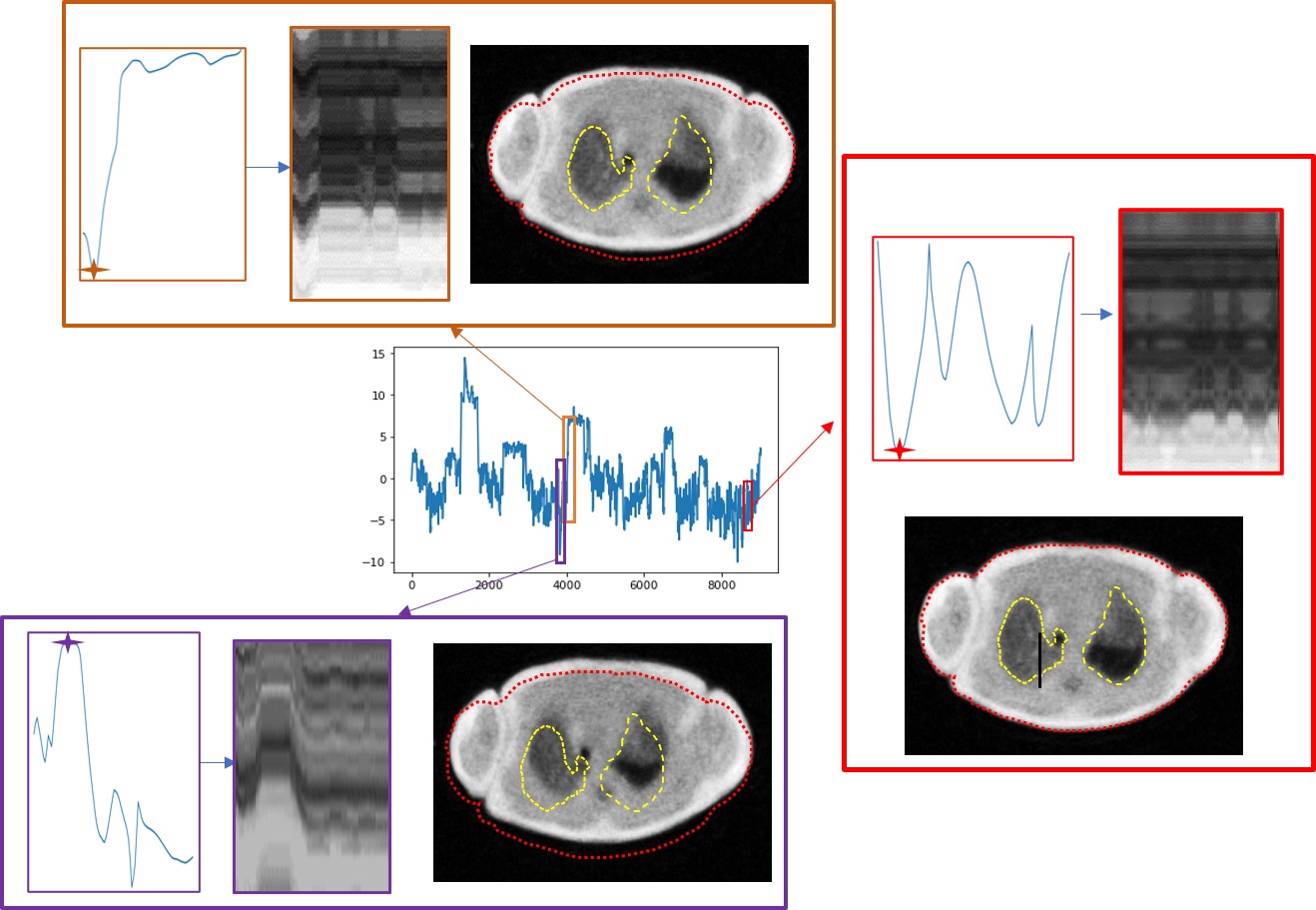}}
	\subfigure[Comparison based on the axial view]{\includegraphics[width=0.65\textwidth]{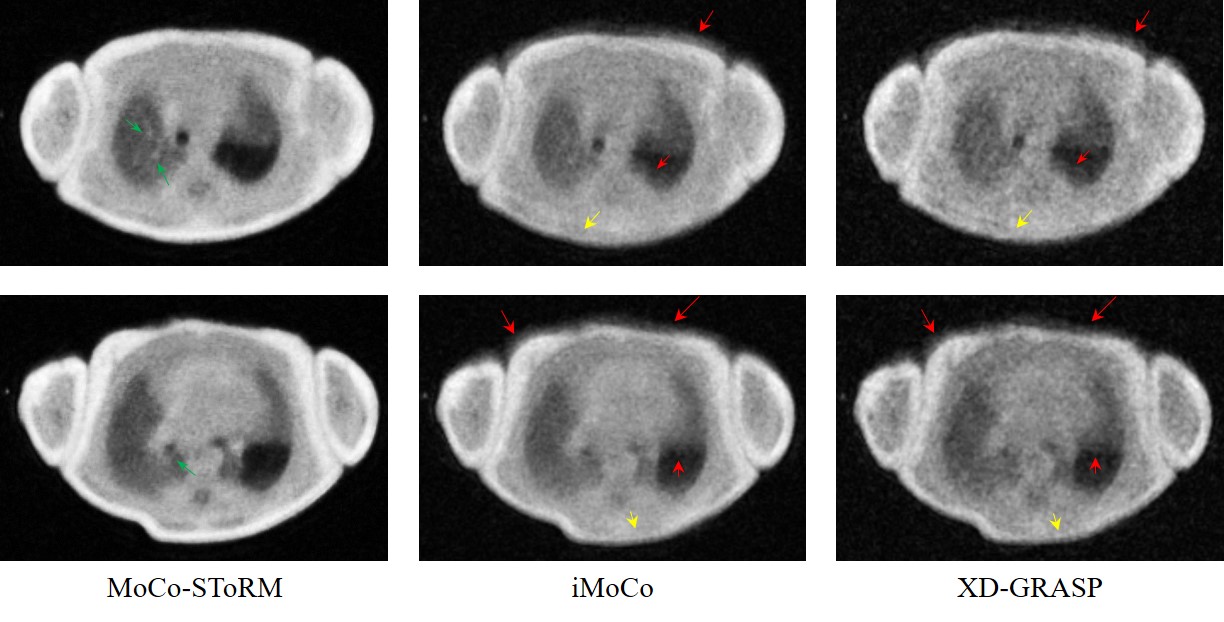}}
    \subfigure[Comparison based on the sagittal view]{\includegraphics[width=0.65\textwidth]{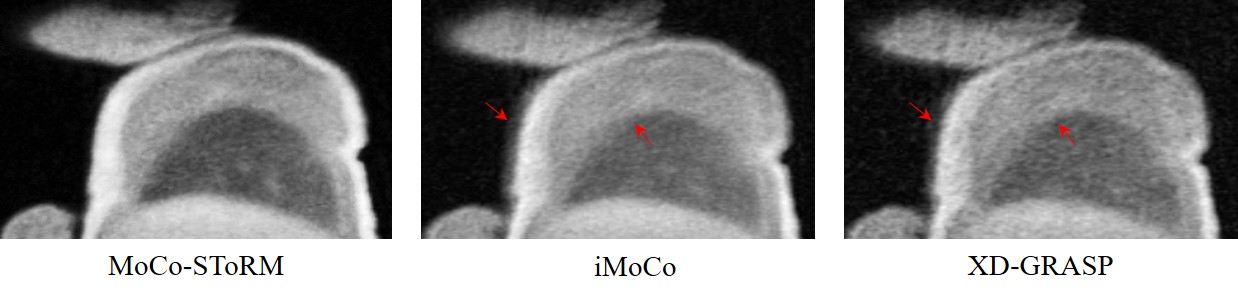}}
	\caption{\small{Study of the neonatal subject in the NICU. (a) shows the latent vectors estimated from the proposed scheme, and three parts are zoomed to study the bulk motions. In (b) and (c), we showed the comparison of the reconstructions from different methods based on the two different views. We can see that iMoCo and XD-GRASP suffered from heavy motion artifacts and failed to capture any details in the lung.}}
	\label{nicu}
\end{figure}

\subsubsection{Feasibility study in neonatal imaging:}

Neonatal subjects often suffer from several developmental lung disorders; the non-ionizing nature of MRI radiation makes it the ideal modality to image the lung of these subjects \cite{higano2017retrospective,hahn2017pulmonary,higano2018neonatal}. A low-field, low-footprint MRI system was considered in \cite{higano2017retrospective,higano2018neonatal}. By imaging the neonatal subjects within the NICU, this approach minimizes the risk of infection. The subject was imaged using a body coil, which offers limited SNR compared to the multi-coil array used in the adult setting. One of the main challenges with neonatal MRI is bulk motion, especially when the subjects are awake. In this work, we study the feasibility of the proposed scheme to offer motion compensation in a challenging subject with extensive bulk motion, which was challenging for the conventional methods. We show the results in Fig. \ref{nicu}. In Fig. \ref{nicu}.(a), we show the motion signal (latent vectors) estimated from the proposed MoCo-SToRM scheme. We note that there are several discontinuities in the latent signal, which correspond to bulk motion events. Two of the bulk motion events are highlighted in the red and purple boxes, respectively. Besides the two examples of bulk motion, we also zoom into a section with no bulk motion. We also show the reconstructed images corresponding to the marked positions in the latent vectors in each of the sub-series in the respective boxes. The bulk motions in the two examples can be clearly seen from the boundary of the body, denoted by the red dotted curve. Furthermore, we see that when the patient was in different positions, the shape of the lung was different, as indicated by the yellow dotted curves. In Fig. \ref{nicu}.(b), we show the comparisons with iMoCo and XD-GRASP based on two slices from the axial view. From the figure, we see that MoCo-SToRM is able to reduce the motion artifacts, as indicated by the red arrows in the figures. Furthermore, we can see that the boundaries in iMoCo and XD-GRASP are blurred out due to bulk motions, as shown by the yellow arrows. However, MoCo-SToRM is able to reconstruct the boundaries. Also, we can see that some details in the lung region are captured in the MoCo-SToRM reconstructions, as indicated by the green arrows. Fig. \ref{nicu}.(c) shows the sagittal view of the comparisons.

\subsection{Maximum intensity projections of the reconstructions}

\begin{figure}[!htbp ]
\centering
           \includegraphics[width=0.8\textwidth]{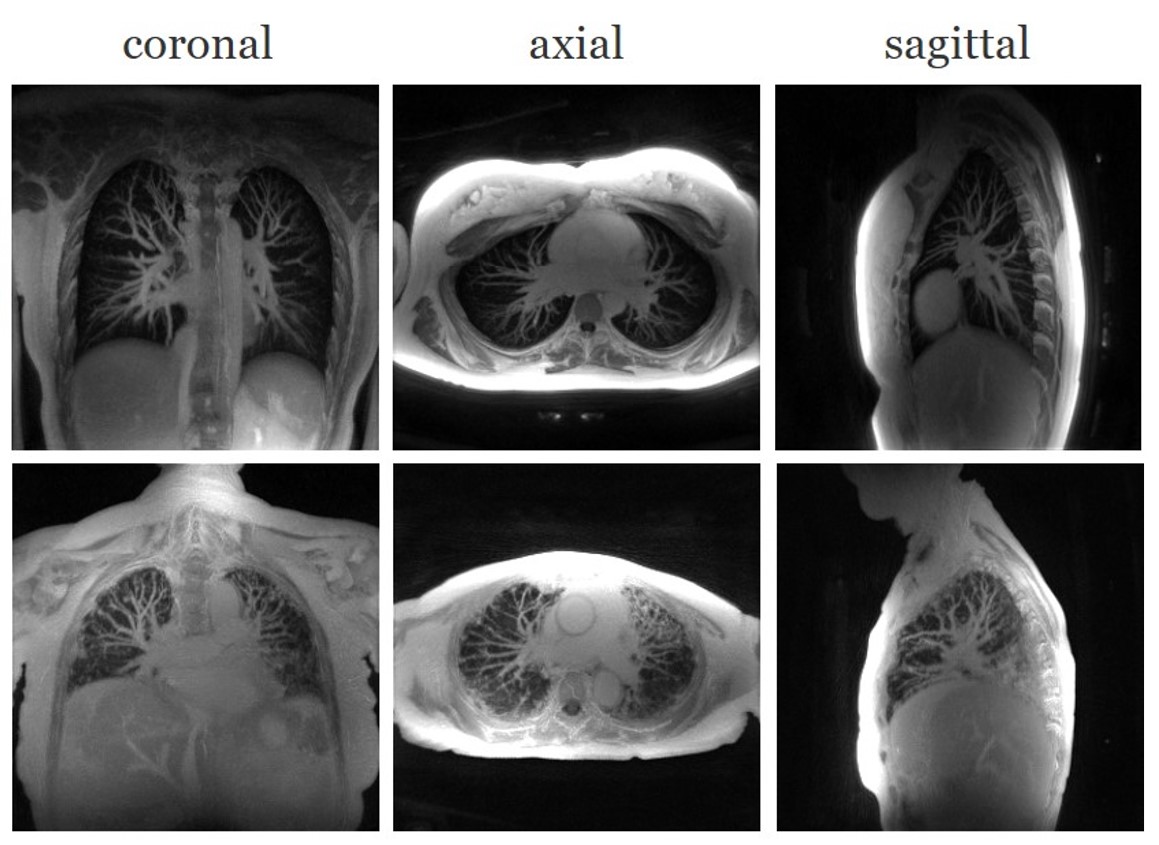}
	\caption{\small{Showcase of the proposed scheme on the post-contrast data from both the healthy subject and the fibrotic subject. Maximum intensity projection of three views are shown for each dataset.}}
	\label{showcase}
\end{figure}

In this section, we show some results that we obtained from the proposed MoCo-SToRM reconstructions. In Fig. \ref{showcase}, we show the reconstructions obtained from two post-contrast datasets, one from a healthy subject and anther from a diseased subject. Maximum intensity projection (MIP) \cite{napel1992ct} is used to generate the results. MIP is known to have the benefit that the vascular structures can be clearly seen as tubular and branching structures in MIP images \cite{gruden2002incremental}. By showing the three views for the reconstruction using MIP, the lung structure and vascular structures for each subject can be seen in a direct way, which can be readily used by doctors in clinics. For each sub-figure in Fig. \ref{showcase}, 20 slices are used for MIP images.

More results, including some movies on the reconstructions and showcases of the bulk motions in both the adult subject and the neonatal subject, can be found on our website: \url{https://sites.google.com/view/qing-zou/blogs/moco-storm}.

\section{Discussion \& Conclusion}

In this work, we proposed an unsupervised motion-compensated scheme using smoothness regularization on manifolds for the reconstruction of high-resolution free-breathing lung MRI. The proposed algorithm jointly estimates the latent vectors that capture the motion dynamics, the corresponding deformation maps, and the reconstructed motion-compensated images from the raw k-t space data of each subject. Unlike current motion-resolved strategies, the proposed scheme is more robust to bulk motion events during the scan, which translates to less blurred reconstructions in datasets with extensive motion. The proposed approach may be applicable to pediatric and neonatal subjects that are often challenging to image using traditional approaches. 

In this study, we restricted our attention to 1-D latent vectors. In our future work, we will consider its extension using higher-dimensional latent space, which will allow improved robustness to different motion components, including cardiac motion and bulk motion. The challenge with the direct extension of the proposed scheme to this setting is the increased computational complexity. 

A difference between the proposed scheme and the motion-resolved reconstruction is that rather than just resolve some phases, the proposed scheme is able to get the temporal-resolved reconstruction with 0.1s temporal resolution. This means that we are able to have more intermediate motion states ($\sim$ 25 states) between the exhalation state and the inhalation state. The proposed MoCo-SToRM scheme is also able to deal with bulk motions as discussed in the previous section. This offers the possibility of using the proposed scheme for patient groups such as pediatric patients.

%\section{Conclusion}

%In this work, we introduced an unsupervised motion-compensated reconstruction scheme called MoCo-SToRM for free‐breathing pulmonary MRI. We assumed that the image frames in the time series were formulated as the deformed version of the 3D template image volume and that the deformation maps were the output of a CNN-based generator with low-dimensional latent inputs. The parameters of the generator, latent vectors, and image template were jointly learned from the undersampled k-t space data. The proposed scheme was first tested using a simulation study. The algorithm was also used for real data acquired from both healthy and diseased subjects. The comparison of the scheme with the state-of-the-art motion-resolved and motion-compensated reconstructions demonstrates the preliminary utility of the proposed approach for motion-compensated free‐breathing pulmonary MRI reconstruction. 

\section*{Acknowledgement}

This work is supported by NIH under Grants R01EB019961, R01AG067078-01A1 and R01HL126771.

\section*{References}

\bibliographystyle{ieeetr}
\bibliography{refs}

\end{document}